\documentclass[aps,prl,twocolumn,superscriptaddress,showpacs]{revtex4-1}

\usepackage{bm}        
\usepackage{color}
\usepackage{gensymb}
\usepackage{epstopdf}

\usepackage{graphicx,amsmath,epsfig}

\newcommand{\ignore}[1]{}
\pdfoutput=1
\begin{document}


\title{Rotational state microwave mixing for laser cooling of complex diatomic molecules}


\author{Mark Yeo}
\email{yeoe@jila.colorado.edu}
\affiliation{JILA, National Institute of Standards and Technology and University of Colorado, and Department of Physics, University of Colorado, Boulder, CO 80309}


\author{Matthew T. Hummon}

\affiliation{JILA, National Institute of Standards and Technology and University of Colorado, and Department of Physics, University of Colorado, Boulder, CO 80309}

\author{Alejandra L. Collopy}
\affiliation{JILA, National Institute of Standards and Technology and University of Colorado, and Department of Physics, University of Colorado, Boulder, CO 80309}

\author{Bo Yan}
\affiliation{JILA, National Institute of Standards and Technology and University of Colorado, and Department of Physics, University of Colorado, Boulder, CO 80309}

\author{Boerge Hemmerling}
\affiliation{ Department of Physics, Harvard University, Cambridge, MA 02138, USA  }
\affiliation{Harvard-MIT Center for Ultracold Atoms, Cambridge, MA 02138, USA}

\author{Eunmi Chae}
\affiliation{ Department of Physics, Harvard University, Cambridge, MA 02138, USA  }
\affiliation{Harvard-MIT Center for Ultracold Atoms, Cambridge, MA 02138, USA}

\author{John M. Doyle}
\affiliation{ Department of Physics, Harvard University, Cambridge, MA 02138, USA  }
\affiliation{Harvard-MIT Center for Ultracold Atoms, Cambridge, MA 02138, USA}

\author{Jun Ye}
\affiliation{JILA, National Institute of Standards and Technology and University of Colorado, and Department of Physics, University of Colorado, Boulder, CO 80309}


\date{\today}

\begin{abstract}
We demonstrate the mixing of rotational states in the ground electronic state using microwave radiation to enhance optical cycling in the molecule yttrium (II) monoxide (YO).  This mixing technique is used in conjunction with a frequency modulated and chirped continuous wave laser to slow longitudinally a cryogenic buffer gas beam of YO.   We generate a measurable flux of YO below 10~m/s, directly loadable into a three-dimensional magneto-optical trap. This technique opens the door for laser cooling of molecules with more complex structure.
\end{abstract}

\pacs{37.10.Mn, 37.10.Pq, 37.10.Vz}

\maketitle

The field of cold and ultracold molecules is undergoing rapid growth with the development of new techniques and applications~\cite{Carr2009Review}. Cold polar molecules are continuing to improve tests of fundamental symmetries~\cite{AcmeEDM2014,HindsEDM2011,Hunter2012,Zhuang2011,Israev2010}. The strong dipolar interaction present in polar molecules also yields a rich set of applications in novel quantum matter~\cite{Buchler2007,Yao2013,Yan2013} and cold chemistry~\cite{Ospelkaus12022010,Bell2009,Sawyer2011}.

In many of these applications it is desirable to have a slow source of molecules to either enhance the interaction time or for loading into a trap.  The magneto-association and coherent state transfer of ultracold alkali atoms~\cite{Ni2008} can produce polar molecules near quantum degeneracy. Supersonic beams produce molecules with forward velocity $\gtrsim$ 300~m/s~\cite{Scoles}, whereas cryogenic buffer gas beams have velocities of 50-200~m/s, depending on source configuration~\cite{hutzler2012buffer} and with the lower velocity beams directly loadable into a trap~\cite{Lu2014}. Many techniques to further slow these beams have been developed. Stark~\cite{VandeMeerakker2006,Sawyer2008} and Zeeman~\cite{Narevicius2008,Hogan2008} deceleration have been used to load conservative traps, which can be used for evaporation~\cite{Stuhl2012}.  Centrifugal slowing~\cite{Chervenkov2014} and opto-electric cooling was demonstrated with CH$_3$F molecules~\cite{Zeppenfeld2012}.  The rovibrational branching that prevents optical cycling transitions in molecules was addressed theoretically~\cite{DiRosa2004,StuhlTiO} and optical Doppler cooling was subsequently demonstrated in the SrF system~\cite{Shuman2010}. Magneto-optical trapping was proposed for TiO~\cite{StuhlTiO} and demonstrated in two dimensions for YO~\cite{Hummon2013} and in three dimensions for SrF~\cite{DeMille3DMOT,McCarron2014}.

While it is possible to load directly a buffer-gas cooled atomic beam into a three-dimensional (3D) magneto-optic trap (MOT)~\cite{Hemmerling2014}, for molecules with much lower photon scattering rates longitudinal optical cooling is necessary for loading into a 3D MOT. Typically $10^4$ photons must be scattered to slow molecules to within the capture velocities of a normal MOT.  During the slowing process, the molecular transition used for optical cooling experiences a Doppler frequency shift of tens or hundreds of resonance linewidths, resulting in inefficient photon scattering for a monochromatic unchirped laser beam. To date, molecules that have been laser cooled all have the magnetic sub-levels in the ground electronic state continually remixed by the multi-leveled optical cycling process.  Hence, the widely used atomic Zeeman slower~\cite{PhillipsZeemanSlower}, which relies on the Zeeman shift of a single magnetic sublevel to compensate for the Doppler shift during deceleration is not applicable.  To maintain a sufficiently large optical scattering rate throughout the slowing process, two techniques have been employed. Chirping the laser frequency was shown to reduce the velocity of a supersonic beam of CaF by 30~m/s~\cite{Zhelyazkova2014}. For SrF, broadband laser radiation was used to slow molecules to be loaded directly into a 3D MOT~\cite{DeMille3DMOT}.

The advent of optical cycling in molecules \cite{StuhlTiO,Shuman2010} is critical for the radiation pressure force to directly slow and cool molecular beams. However, despite the achievement of quasi-closed optical cycling in several molecular systems, preventing optical pumping into dark states still remains difficult for a large class of molecules.  While vibrational dark states can be addressed with repump lasers \cite{DiRosa2004},  rotational dark states often have parity selection rules that prevent direct optical repumping.  There are several loss mechanisms that break the rotational closure in the current schemes for molecular optical cycling transitions, and they arise from additional decays via intermediate states or higher order transition moments.  Vibrational states typically have lifetimes $\sim$1~s \cite{Vanhaecke2007}, and rovibrational decay is not yet a limiting factor for current techniques.  Decays via magnetic dipole transitions are typically suppressed by a factor of $\alpha^2\approx5\times 10^{-5}$ compared to electric dipole transitions ($\alpha$ is the fine structure constant), though recent measurements in OH indicate that they can sometimes be a factor of 10 faster than expected \cite{Kirste2012}. Decays through intermediate electronic states, present in for example YO \cite{Chalek1976} and BaF \cite{Allouche93}, are suppressed by the relatively long transition wavelengths. For YO, the suppression factor is $\sim$$3\times10^{-4}$, however the decay is sufficient to cause population leakage to rotational dark states that reduce photon scattering.  In this Letter, we report microwave mixing of rotational states to close these additional loss channels. This enables us to slow longitudinally a beam of YO emitting from a two-stage buffer gas cell to molecular velocities $<$ 10~m/s, which can be loaded into a 3D MOT.

\begin{figure}[t]
\includegraphics[width=3.4 in]{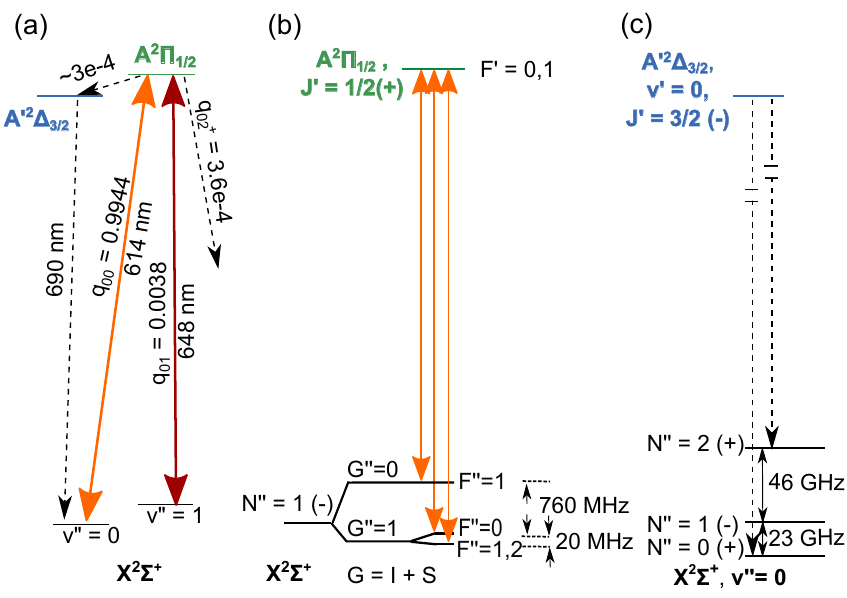}
\caption[YO Level Structure]{(Color online) (a) YO vibronic structure.  Dashed arrows indicate decay paths with corresponding Franck-Condon factors, $q$ \cite{Bernard1983}.  Solid arrows indicate cooling and repump laser transitions. (b) Hyperfine structure of the $X^2\Sigma$, $A^2\Pi_{1/2}$ states. The solid arrows indicate the three hyperfine pumping components used in each vibrational manifold.  (c) Schematic of rotational mixing.  Dashed arrows indicate the decay paths to $X^2\Sigma$ through the $A'^2\Delta_{3/2}$ state and black solid arrows show the microwave rotational mixing.}\label{fig:YOlevels}
\end{figure}

The optical cycling transition in YO proceeds between the $X^2\Sigma\rightarrow A^2\Pi_{1/2}$ states as described in Ref.~\cite{Hummon2013}.  In brief, highly diagonal Franck-Condon factors limit vibrational branching \cite{Bernard1983} such that a single repump laser limits branching loss to $3.6\times 10^{-4}$ (Fig. \ref{fig:YOlevels}(a)).  Rotational branching is suppressed by a combination of parity and angular momentum selection rules \cite{StuhlTiO,shuman:223001} when driving the $X^2\Sigma,N''=1$ level, where $N$ is the rotation quantum number.  Figure \ref{fig:YOlevels}(b) shows the hyperfine structure of the $X^2\Sigma$ states, where $G$ is the coupled nuclear and electronic spin, and $F$ is the total angular momentum. For the excited $A^2\Pi_{1/2}$ and $A'^2\Delta_{3/2}$ states, $J$ is the total angular momentum excluding nuclear spin. For all three electronic states, $p = \pm$ represents the parity.

Figure \ref{fig:YOlevels}(a) shows the additional loss channel where the $A^2\Pi_{1/2}$, $J'=1/2$ state decays to the $A'^2\Delta_{3/2}$, $J'=3/2$ state.  Decays to states of higher $J$ are forbidden due to angular momentum selection rules.   This state has a radiative lifetime of $\sim$1~$\mu$s \cite{Chalek1976} and will rapidly decay back to the $X^2\Sigma$ state.  Since this is a three-photon process back to the ground state, parity selection rules will only allow decays to even rotational states (Fig. \ref{fig:YOlevels}(c)).  In the $X^2\Sigma$ state, while $J$ is not a good quantum number, each state of $N''>0$ can be expressed as a superposition of states with $J =N \pm 1/2$ (for $N'' = 0$,  J = 1/2).  The $\Delta J=0,\pm1$ angular momentum selection rule prevents decay to $N''\ge 4$ and thus leads to optical pumping into the $N'' = 0, 2$ states.  The rotational constant, $B$, for the $X^2\Sigma$ state is 11.634~GHz.  Microwave radiation tuned to $2B$ and $4B$ can mix the $N'' = 0,2$ levels respectively with the $N''=1$ level.  This mixing effectively removes the rotational dark states; however it also lowers the maximum optical scattering rate as the number of states in $X^2\Sigma$ involved in optical cycling increases by a factor of two.

\begin{figure}[t]
 \includegraphics[width=3.4in]{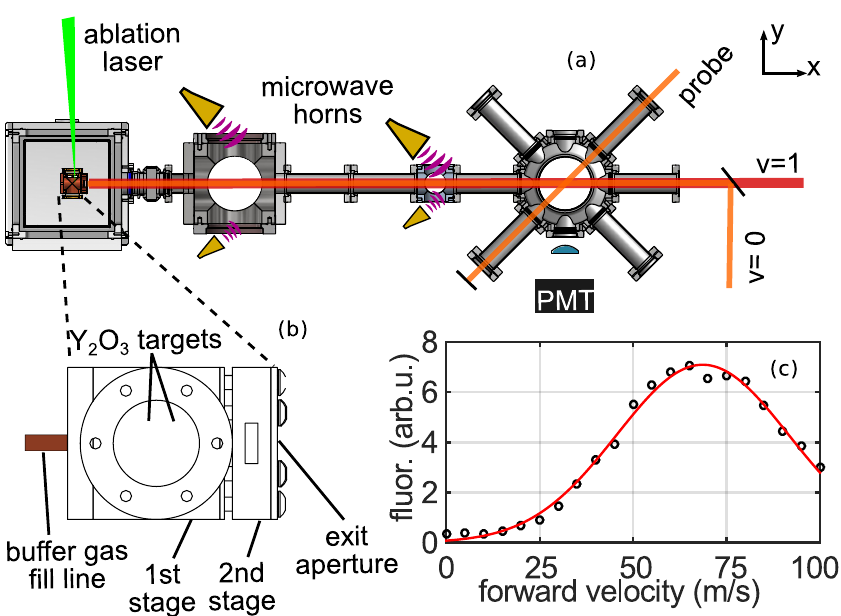}%
 \caption[Apparatus]{(Color online) (a) Depiction of the beam apparatus.  The YO molecular beam interacts with a counter-propagating slowing laser beam along x. A $45\degree$ probe beam is used for Doppler sensitive detection.  A PMT is used to collect fluorescence induced by either the slowing or probe beam.  (b) Two-stage buffer gas cell. (c) Velocity distribution measured at the detection region.  The solid line is a Gaussian fit with center velocity of $69$~m/s and full width at half maximum (FWHM) of $54$~m/s, corresponding to a temperature of 6.7~K.}
\label{fig:App}
 \end{figure}

We employ this rotational state microwave mixing scheme for longitudinal slowing of a YO beam generated from a two-stage cryogenic buffer gas beam source~\cite{hutzler2012buffer}.  We laser ablate (with a 10~Hz repetition rate) sintered Y$_2$O$_3$ pellets in a copper cell filled with helium buffer gas at 3.5~K (Fig. \ref{fig:App}(a)).  The hot YO thermalizes rotationally and translationally to the cold helium buffer gas and is hydrodynamically extracted through a 3~mm aperture into the second stage cell.  By virtue of the 4.5~mm gap between the two stages and the larger exit aperture diameter of $9$ mm, the helium density is $\sim 10$ times smaller in the second stage than that of the first stage and the molecules experience only a few additional collisions before exiting the second stage cell.  This slows the $120$~m/s beam from the first stage~\cite{Hummon2013} to $70$~m/s (Fig.~\ref{fig:App}(c)).  The YO beam then travels $89$~cm to the detection region, during which it may be illuminated by a 6 mm diameter, counter-propagating slowing beam.  The $614$ nm $v'' = 0$ laser has 70~mW of power and the $648$ nm $v'' = 1$ laser has $80$ mW (Fig.~\ref{fig:YOlevels}(a)). The slowing laser has three frequency components to address the hyperfine manifold (Fig.~\ref{fig:YOlevels}(b)).  To provide the rotational mixing, we use two pairs of microwave horns strategically located along the beam path, each pair producing $\sim$1~mW of either the 23~GHz or 46~GHz microwaves. \ignore{The light for the $614$ nm is generated via a frequency doubled fiber based raman amplifier, seeded with an External Cavity Laser Diode (ECDL) and the repump radiation is generated from an ECDL seeded tapered amplifier.}

For velocity sensitive detection, we measure the fluorescence induced by a low intensity 4 mW/cm$^{2}$, 614 nm beam aligned $45\degree$ to the molecular beam. This probe laser has a saturation parameter of $\sim$0.5 and two frequency components separated by 780~MHz (Fig.~\ref{fig:YOlevels}(b)). This eliminates mechanical and optical pumping effects from the detection beam.  The detection laser can be scanned over a range of $\sim$100~MHz at a rate of 1 kHz, corresponding to a velocity detection range of $\sim$100~m/s.

\begin{figure}[t]
 \includegraphics[width=3.4in]{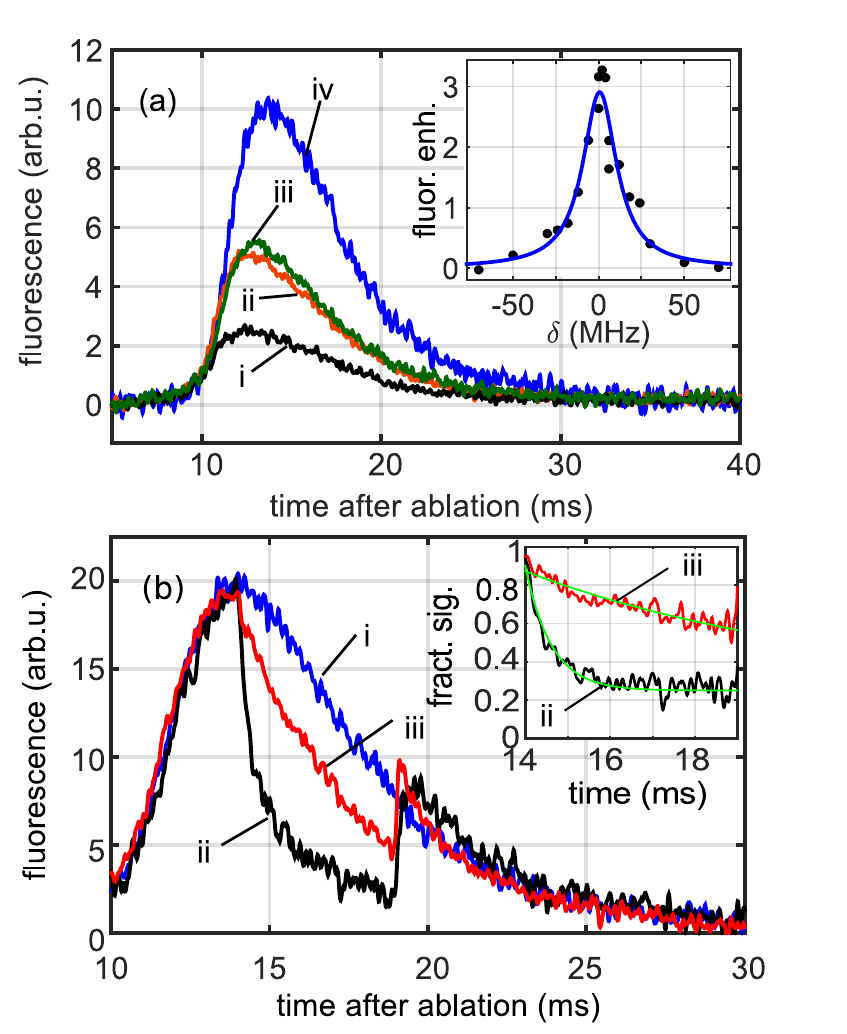}
 \caption[Scattering rate, microwave enhancement]{(Color online) (a) Optical cycling enhancement due to rotational mixing.  The fluorescence is induced by the high power optical slowing beam and $v'' = 1$ repump: (i) No mixing;  (ii) $N''=0\leftrightarrow N'' = 1$ mixing only; (iii) $N''=2\leftrightarrow N'' = 1$ mixing only; (iv) simultaneous $N''=2\leftrightarrow N'' = 1$ and $N''=0\leftrightarrow N'' = 1$ mixing.  The inset shows the fractional enhancement of optical scattering vs. microwave detuning from the rotational resonance.  (b) Comparison of branching losses between $v'' = 1$ and $A'^2\Delta_{3/2}$: (i) Microwaves and $v''=0,1$ are on throughout;  (ii) $v'' = 1$ repump is shut off from 14-19 ms after ablation;  (iii) Microwave mixing is switched off from 14-19 ms. The inset shows curves (ii) and (iii) divided by curve (i).  Solid green lines are exponential fits.}
\label{fig:MWCompare}
 \end{figure}

We study the effects of rotational mixing by observing the fluorescence induced by the slowing beam in Fig.~\ref{fig:MWCompare}(a).  We apply the microwaves in four binary combinations.  For case i, no microwaves are applied. In contrast, when microwave mixing is applied to both $N'' = 0,2$, (curve iv), there is a factor of four increase in fluorescence.   Applying only the $N''=0\leftrightarrow N'' = 1$ or $N''=2\leftrightarrow N'' = 1$ mixing (Fig.~\ref{fig:MWCompare}(b) ii, iii respectively), we see a factor of two increase in fluorescence compared to the no mixing case.  We estimate the vibrational lifetime of the $v''=1$ state from molecular parameters calculated in Ref.~\cite{Langhoff1988} to be $\sim$600~ms, which is much longer than the optical pumping time scale.  The branching ratios for magnetic dipole transitions from the  $A^2\Pi_{1/2}$ state is 2/3(1/3) for the $N'' = 0(2)$ levels in the $X^{2}\Sigma$ state, while electric dipole transitions from the $A'^2\Delta_{3/2}$ state have equal branching ratios to the $N''=0,2$ rotational levels, consistent with curves ii and iii.  Hence, decays through the intermediate $A'^2\Delta_{3/2}$ state are identified to be the dominant process.

To verify that the mixing microwaves are sufficiently strong to address all hyperfine states without the need of additional frequency sidebands, we simultaneously vary the frequency of the $N'' = 1\leftrightarrow N'' = 0 (2)$ mixing microwave by $\delta (2\delta)$ and measure the corresponding enhancement in fluorescence (Fig.~\ref{fig:MWCompare}(a) inset). This enhancement follows a Lorentzian lineshape with a FWHM of $24$~MHz, which is larger than the hyperfine variations of $<$5~MHz among the three relevant rotational manifolds~\cite{Childs1988,BrownAndCarrington}. The mixing for higher vibrational states is suppressed because the rotational constant $B$ for $v'' = 1$ is $50$~MHz smaller than that for $v'' = 0$.  This results in a $100$ $(200)$~MHz detuning for the $N'' =0$ $(2)$ transition in $v'' = 1$. Hence, we do not suffer any further reduction in the maximum optical cycling rate.

Figure \ref{fig:MWCompare}(b) shows a measurement of the branching loss from the $A^2\Pi_{1/2}$ state to the $A'^2\Delta_{3/2}$ state.  Again we measure the fluorescence induced by the high power slowing beam. For case i, the slowing and repumping light as well as microwaves are turned on at all times.  For case ii, the $v'' = 1$ repump laser is rapidly shut off in $\sim$10~$\mu$s from $14$ ms to 19 ms after ablation.  This leads to an abrupt decrease in the fluorescence as the YO molecules are rapidly pumped into the $v'' = 1$ state and hence turn dark. To extract this pumping rate, we divide curve (ii) by curve (i) and fit an exponential with a constant background ratio of 0.25, resulting in a $1/e$ time constant of 620~$\mu$s.  Combining this with the calculated Franck-Condon factor $q_{01} =3.8\times10^{-3}$ \cite{Bernard1983}, yields a photon scattering rate of $4.3\times 10^5$~s$^{-1}$.  This is smaller than the maximum possible scattering rate by a factor of $\sim$5~\cite{Hummon2013} and may be due to inefficient depletion due to highly Doppler detuned molecules, insufficient power in the wings of the slowing beam or an overestimate of the $q_{01}$ Franck-Condon factor. Curve iii shows the fluorescence when only the rotational mixing microwaves are turned off between $14-19$~ms.  The $1/e$ decay due to pumping into $N''=0$ and 2 dark states has a $7.4$~ms time constant (inset curve iii). By combining the ratio of the two loss rates and $q_{01}$, we estimate the branching ratio from $A^2\Pi_{1/2}$ to $A'^2\Delta_{3/2}$ to be $\sim$$3\times10^{-4}$.  This branching ratio agrees with the value derived from the transition dipole moments calculated in Ref.~\cite{Langhoff1988} and is similar to the combined branching ratio to $v''\ge2$. At 19 ms after ablation, the vibration repump (curve ii) and mixing microwaves (curve iii) are turned back on and the fluorescence immediately recovers.


   \begin{figure}[t]
 \includegraphics[width=3.4in]{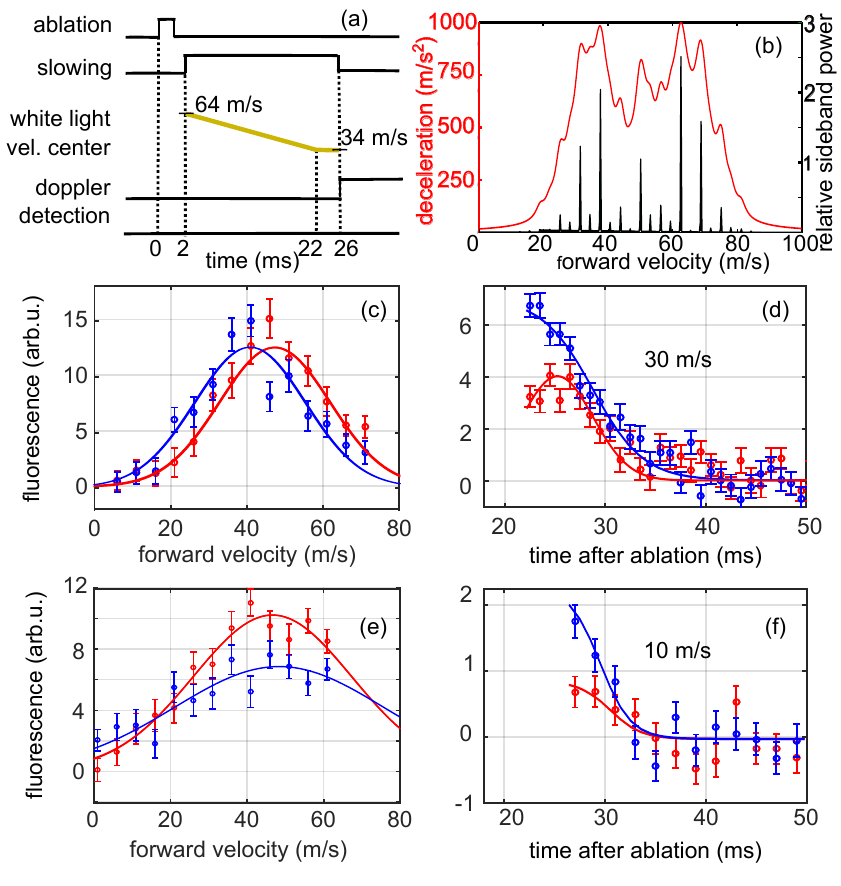}%
 \caption[Slowing]{(Color online) (a) Timing sequence used to slow molecules to below 10~m/s. (b) Example of spectrally broadened slowing light corresponding to a center velocity of 50~m/s.  The black lines shows the sidebands generated from an EOM and rapid polarization switching. The red curve shows molecular deceleration with an optical saturation parameter of $3$. (c-f) Comparisons between no slowing in red and with slowing in blue.  Solid lines are guides to the eye. (c) Doppler spectra for a 10~ms long slowing sequence.  (d) Time of flight for $30$~m/s molecules.  (e) Doppler spectra for a $24$ ms long slowing sequence as depicted in (b). (f) Time of flight for $10$~m/s molecules.}
\label{fig:Slowing}
 \end{figure}

We estimate the capture velocity of a 3D MOT for YO with a laser beam diameter of 1 cm to be $10$~m/s.  No detectable population of 10~m/s molecules exists in the buffer gas beam source. To slow some fraction of the molecules to below this capture velocity, we apply longitudinal radiation pressure slowing with protocols shown in Fig.~\ref{fig:Slowing}(a).  To increase the fraction of molecules addressed by the slowing beam, we spectrally broaden the light~\cite{Zhu1991} via a resonant electro-optic modulator (EOM) with a modulation frequency of $10$~MHz and a phase modulation index of $3.3$.  This spectrally broadens each laser frequency to $\pm 30$~MHz, corresponding to a velocity spread of $\pm20$~m/s. Polarization switching used for destabilizing Zeeman dark states~\cite{Hummon2013} further adds $5$~MHz sidebands and thus we obtain frequency components spaced by the natural linewidth of the main cooling transition, ensuring that we have continuous deceleration across this velocity range (Fig.~\ref{fig:Slowing}(b)).

As the range of the slowing force is smaller than the velocity spread of the YO beam itself, we additionally implement frequency chirping on the slowing light.  In Fig.~\ref{fig:Slowing}(c) and (d), the slowing sequence is designed to enhance molecules at $30$~m/s.  The molecules propagate freely for the first 12 ms after ablation, then the slowing beam is turned on from 12-22~ms.  During this time, the center of the broadband radiation is linearly swept from being resonant with $64$~m/s to $44$~m/s molecules. In panel (c), we see a molecular population enhancement between $20-40$~m/s (negligible for below $20$~m/s), and depletion for velocities greater than 40~m/s. In panel (d), we maintain the probe laser's frequency to be resonant with the $30$~m/s molecule and we observe a clear enhancement in the number of molecules in this velocity class compared to the original beam. Due to off-resonant excitations, some faster molecules ($35$~m/s) contribute to the detected signal at 27 ms for the unslowed case.

To produce even slower molecules, we increase the time during which the slowing laser is turned on and perform a frequency chirp to address lower velocity molecules. The slowing light is turned on 2~ms after ablation and is left on for the next 24~ms. During the first $20$~ms, the center frequency of the broadband slowing laser is linearly increased from being resonant with 64~m/s to 34~m/s molecules (Fig.~\ref{fig:Slowing}(a)).  Figure~\ref{fig:Slowing}(e) shows the velocity distribution after the slowing sequence compared to the original velocity distribution. Slowing enhances the molecular population below 20~m/s in the detection region and depletes molecules with velocities higher than 20~m/s. Again, we tune the Doppler-sensitive probe laser to be resonant with the 10~m/s molecules and we see a clear enhancement in the number of molecules in this velocity class (Fig.~\ref{fig:Slowing}(f)).  Unslowed molecules leaving the cell with 10~m/s forward velocity will arrive 89~ms after the ablation, well after the detection time.  Hence, we attribute the non-zero 10~m/s signal in the unslowed case to off-resonant excitations from fast molecules.

We have demonstrated enhanced optical cycling in YO via microwave mixing of rotational states in the ground electronic state.  This enabled us to slow longitudinally the YO beam via radiation pressure, and hence produce molecules that can be loaded directly into a 3D MOT.  To further close the vibrational branching to $<10^{-6}$, an additional $v'=1\leftarrow v''=2$ repump laser at $649$ nm can be used~\cite{Hummon2013,Bernard1983}.  By turning off the microwave mixing for $N''=1\leftrightarrow N''=0$ at an appropriate time, we can accumulate trapped cold YO in the ro-vibrational ground state.  The $A'^2\Delta_{3/2}$ state lifetime is $\sim$1~$\mu$s, which is $>$10 times longer than that of $A^2\Pi_{1/2}$. This opens up the possibility of a narrow line MOT~\cite{Loftus2004,Collopy2015} with a Doppler temperature of $10$~$\mu$K as compared to the $116$ $\mu$K obtainable from the $X^2\Sigma$, $v'=0$ to $A^2\Pi_{1/2}$ transition.
\begin{acknowledgments}
We thank M. Petzold and M. Kuhnert for their contributions to the early stage of this work. We acknowledge funding support from ARO and AFOSR (MURI), Gordon and Betty Moore Foundation, NIST, and NSF.
\end{acknowledgments}

%

\end{document}